\documentclass[12pt]{article}
\usepackage{graphicx}

\begin{document}

\title{A Statistical Approach to Describe Highly Excited Heavy and Superheavy Nuclei }

\author{Peng-Hui Chen$^{1,2}$, Zhao-Qing Feng$^{2,3}$  \footnote{Corresponding author. Tel. +86 931 4969152. \newline \emph{E-mail address:} fengzhq@impcas.ac.cn (Z.-Q. Feng)}, Jun-Qing Li$^{2}$, Hong-Fei Zhang$^{1}$}

\date{}
\maketitle

\begin{center}
$^{1}${\small \emph{School of Nuclear Science and Technology, Lanzhou University, Lanzhou 730000, People's Republic of China }}           \\
$^{2}${\small \emph{Institute of Modern Physics, Chinese Academy of Sciences, Lanzhou 730000, People's Republic of China }}           \\
$^{3}${\small \emph{Kavli Institute for Theoretical Physics, Chinese Academy of Sciences, Beijing 100190, People's Republic of China }}
\end{center}

\textbf{Abstract}
\par
A statistical approach based on the Weisskopf evaporation theory has been developed to describe the de-excitation process of highly excited heavy and superheavy nuclei, in particular for the proton-rich nuclei. The excited nucleus is cooled by evaporating $\gamma$-ray, light particles (neutrons, protons, $\alpha$ etc) in competition with the binary fission, in which the structure effects (shell correction, fission barrier, particle separation energy) contribute to the processes. The formation of residual nuclei is evaluated via sequential emission of possible particles above the separation energies. The available data of fusion-evaporation excitation functions in the $^{28}$Si+$^{198}$Pt reaction can be reproduced nicely well within the approach.
\newline
\emph{PACS}: 25.70.Jj, 24.10.-i, 25.60.Pj    \\
\emph{Keywords:}  Excited nucleus; Structure effects; Weisskopf evaporation theory; Fusion-evaporation reactions

\bigskip

The synthesis of very heavy (superheavy) nuclei is a very important subject in nuclear physics motivated by reaching the island of stability predicted theoretically, which has obtained much progress in experiments with fusion-evaporation reaction mechanism\cite{Ho00,Mo04,Og07}. The existence of superheavy nucleus (SHN) ($Z\geq106$) is due to a strong shell effect against the large Coulomb repulsion. However, the shell effect will be reduced by the increasing excitation energy, which leads to the decrease of the survival of residue nucleus. The fusion-evaporation reaction to form superheavy compound nucleus can be understood as three stages in accordance with the evolution of two heavy colliding nuclei, namely the capture process of the colliding system to overcome Coulomb barrier, the formation of the compound nucleus to pass over inner fusion barrier, as well as the de-excitation of the thermal compound nucleus against fission\cite{Fe06}. The survival probability as the last stage is particularly important in the evaluation of production cross sections of heavy and superheavy nuclei. Besides the synthesis of SHN, one can produce proton-rich nuclei close to the proton drip line using the fusion-evaporation reactions.

 In this letter, we extend the statistical approach based on the Weisskopf evaporation theory\cite{We37} to describe the de-excitation process of highly excited proton-rich heavy nuclei. Evaporation of light charged particles in the statistical model has been implemented. It is well known that the evaporation residue cross section in the fusion reactions is evaluated as a sum over partial angular momentum $J$ at the centre-of-mass energy $E_{c.m.}$  in the evaporation channel $s$,
\begin{eqnarray}
\sigma^s_{ER}(E_{c.m.}) =&&\frac{\pi\hbar^2}{2\mu E_{c.m.}} \sum \limits ^{J_{max}}_{J=0} (2J+1)T(E_{c.m.},J)P_{CN}(E_{c.m.},J)       \nonumber  \\
&&  \times   W^s_{sur}(E_{c.m.},J)
\end{eqnarray}
Here, $T(E_{c.m.})$ is the penetration probability of the two colliding nuclei overcoming the Coulomb barrier to
form the DNS, which is calculated using the empirical coupled channel model\cite{Fe06,Za03}. The $P_{CN}$ is the probability that the heavy system evolves from a touching configuration into the formation of compound nucleus in competition with quasi-fission and fission of the heavy fragments. However, for the light reaction systems or projectile-target combinations with larger mass asymmetry ($A_{P}/A_{T}<0.1$) the probability is close to unity, $P_{CN}\approx 1$ \cite{Ya13}.

The survival probability is particularly important in evaluation of the cross section, which is usually calculated with the statistical approach. The physical process in understanding the excited nucleus is clear. But the magnitude strongly depends on the ingredients in the statistical model, such as level density, separation energy, shell correction, fission barrier etc. The excited nucleus is cooled by evaporating $\gamma$-ray, light particles (neutrons, protons, $\alpha$ etc) in competition with fission. Similar to neutron evaporation solely\cite{Fe06}, the probability in the channel of evaporating the $x-$th neutron, the $y-$th proton and the $z-$ alpha is expressed as
\begin{eqnarray}
W_{sur}(E^*_{CN},x,y,z,J) =  P(E^*_{CN},x,y,z,J)   \times   \prod^x_{i=1} \frac{\Gamma_n(E^*_i,J)}{\Gamma_{tot}(E^*_i,J)} \times  
\prod^y_{j=1} \frac{\Gamma_p(E^*_j,J)}{\Gamma_{tot}(E^*_j,J)}   \times   \prod^z_{k=1} \frac{\Gamma_{\alpha}(E^*_k,J)}{\Gamma_{tot}(E^*_k,J)}.
\end{eqnarray}
Here the $E^*_{CN}$, $J$ are the excitation energy and the spin of the excited nucleus, respectively. The total width $\Gamma_{tot}$  is the sum of partial widths of particle evaporation, $\gamma$-emission and fission. The excitation energy $E^*_s$ before
evaporating the $s$-th particle is evaluated by
\begin{equation}
E^*_{s+1} = E^*_s - B^n_i - B^p_j - B^{\alpha}_k - 2T_s
\end{equation}
with the initial condition $E^*_1 = E^*_{CN}$ and $s=i+j+k$. The $B^n_i$, $B^p_j$, $B^{\alpha}_k$ are the separation energy of the $i$-th neutron, $j$-th proton, $k$-th alpha,respectively. The nuclear temperature $T_i$ is given by $E^*_i = aT_i^2-T_i$ with $a$ being the level density parameter.

Assuming the electric dipole radiation(L=1) dominating $\gamma-$emission, the decay width is calculated by
\begin{equation}
\Gamma_\gamma(E^*_{CN},J) = \frac{3}{\rho(E^*,J)}    \nonumber  \\
\int \limits ^{E^*-\delta -\frac{1}{a}}_{\varepsilon = 0} \rho(E^*-E_{rot}-\varepsilon,J) f_{E_1}(\varepsilon) d \varepsilon,
\end{equation}
and
\begin{equation}
f_{E_1}(\varepsilon) = \frac{4}{3\pi} \frac{1+\kappa}{mc^2} \frac{e^2}{\hbar c} \frac{NZ}{A} \frac{\Gamma_G\varepsilon^4}{(\Gamma_G\varepsilon)^2 + (\Gamma_G^2 - \varepsilon^2)^2}.
\end{equation}
Here, $\kappa = 0.75$, $\Gamma _G$ and $E_G$ are width and position of the electric dipole resonance, for heavy nucleus ,$\Gamma _G = 5$ MeV\cite{Sc91},
\begin{equation}
E_G = \frac{167.23}{A^{1/3}\sqrt{1.959 + 14.074A^{-1/3}}}.
\end{equation}

The particle decay widths are evaluated with the Weisskopf evaporation theory as \cite{We37}
\begin{eqnarray}
\Gamma_\nu(E^*,J) = (2s_\nu + 1) \frac{m_\nu}{\pi^2 \hbar^2 \rho(E^*,J)} \int \limits ^{E^*-B_\nu-\delta-\delta_n -\frac{1}{a}}_0
\nonumber\\ 
\varepsilon \rho(E^*-B_\nu - \delta_n-E_{rot}-\varepsilon,J)\sigma_{inv}(\varepsilon) d \varepsilon .
\end{eqnarray}
Here, $s_\nu$, $m_\nu$ and $B_\nu$  are the spin, mass and binding energy of evaporating particle, respectively. The pairing correction energy $\delta$ is set to be $12/\sqrt{A}, 0, -12/\sqrt{A}$  for even¨Ceven, even¨Codd and odd¨Codd nuclei, respectively. The inverse cross section is given by $\sigma_{inv}=\pi R_\nu^{2}T(\nu) $. The penetration probability is set to be unity for neutron and $T(\nu) =(1 + \exp(\pi(V_C(\nu)-\varepsilon)/\hbar\omega))^{-1}$ for charge particle with $\hbar \omega= 5 $ and 8 MeV for proton and alpha, respectively. The fission width is calculated with the similar method in Ref. \cite{Fe07,Fe09}.

\begin{figure}
\begin{center}
{\includegraphics*[width=1.\textwidth]{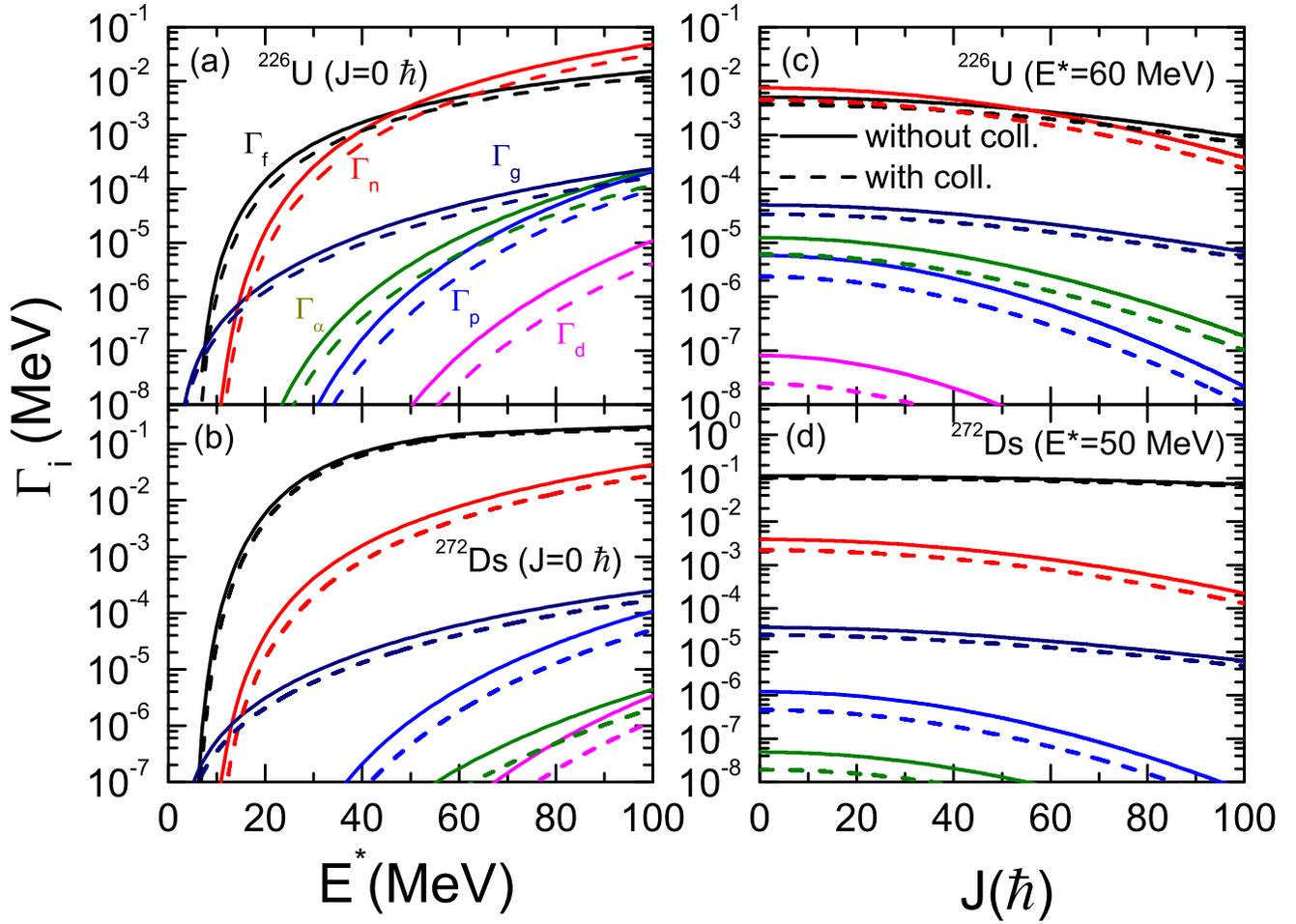}}
\end{center}
\caption{Partial decay widths of proton-rich nucleus $^{226}$U and superheavy nucleus $^{272}$Ds as functions of excitation energies and angular momenta.}
\end{figure}

The level density is calculated from the Fermi-gas model\cite{Ig79} as,
\begin{equation}
\rho(E^*,J) = K_{coll}\cdot \frac{2J+1}{24\sqrt{2}\sigma^3a^{1/4}(E^*-\delta)^{5/4}}  \nonumber\\[1mm]
exp\left[ 2\sqrt{a(E^*-\delta)} - \frac{(J+1/2)^2}{2\sigma^2}\right],
\end{equation}
with $\sigma^2 = 6\bar{m}^2\sqrt{a(E^*-\delta)}/\pi^2$  and $\bar{m}\approx0.24A^{2/3}$. The $K_{coll}$ is the collective enhancement factor, which includes the rotational and vibrational effects \cite{Fe09,Ju98}. The level density parameter is related to the shell correction $\mathbf{energy}$ $E_{sh}(Z,N)$ and the excitation energy $E^{\ast}$ of the nucleus as
\begin{equation}
a(E^{\ast},Z,N)=\tilde{a}(A)[1+E_{sh}(Z,N)f(E^{\ast}-\Delta)/(E^{\ast}-\Delta)].
\end{equation}
Here, $\tilde{a}(A)=\alpha A + \beta  A^{2/3}b_{s}$ is the asymptotic Fermi-gas value of the level density parameter at high excitation energy. The shell damping factor is given by
\begin{equation}
f(E^{\ast})=1-\exp(-\gamma E^{\ast})
\end{equation}
with $\gamma=\tilde{a}/(\epsilon  A^{4/3})$. The parameters $\alpha$, $\beta$, $b_{s}$ and $\epsilon$ are taken to be 0.114, 0.098, 1. and 0.4, respectively \cite{Fe09}.
Shown in Fig. 1 is a comparison of the partial decay width for proton-rich nucleus $^{226}$U and superheavy nucleus $^{272}$Ds. It is obvious that the charged particles (p, $\alpha$) have smaller widths for the superheavy nucleus in comparison to  the proton-rich nucleus because of larger separation energies. The fission width increases rapidly in the excitation energy range of 10-30 MeV for superheavy nucleus and the larger width leads to a smaller survival probability, which is caused from the fact that the fission barrier decreases exponentially with increasing the excitation energy \cite{Fe06}. The collective enhancement factor increases the level density, but reduces the partial widths, in particular for particle evaporation.

For one particle evaporation, the realization probability is given by
\begin{equation}
P(E^*_{CN},J) = \exp\left( -\frac{(E^*_{CN} - B_s - 2T)^2}{2\sigma^2} \right).
\end{equation}
The width $\sigma$ is taken to fit the experimental width of fusion-evaporation excitation functions. The realization probability $P(E^*_{CN},x,y,z,J)$ for evaporating the $x$ neutrons, $y$ protons, $z$ alphas at the excitation energy of $E^*_{CN}$ and angular momentum of $J$ is calculated by the Jackson formula\cite{Ja56} as
\begin{equation}
P(E^*_{CN},s,J) = I(\Delta_s,2s-3) - I(\Delta_{s+1},2s-1),
\end{equation}
where the quantities $I$ and $\Delta$ are given by following:
\begin{equation}
I(z,m) = \frac{1}{m!} \int ^z_0 u^m e^{-u} d u,   \\
\Delta_s = \frac{E^*_{CN} - \sum \limits ^s_{i=1}B^\nu_i}{T_i}.
\end{equation}
The $B^\nu_i$ is the separation energy of evaporating the $i$-th particle and $s(x,y,z)=x+y+z$. The spectrum of the realization probability determines the structure of survival probability in each evaporation channel.

\begin{figure}
\begin{center}
{\includegraphics*[width=1.\textwidth]{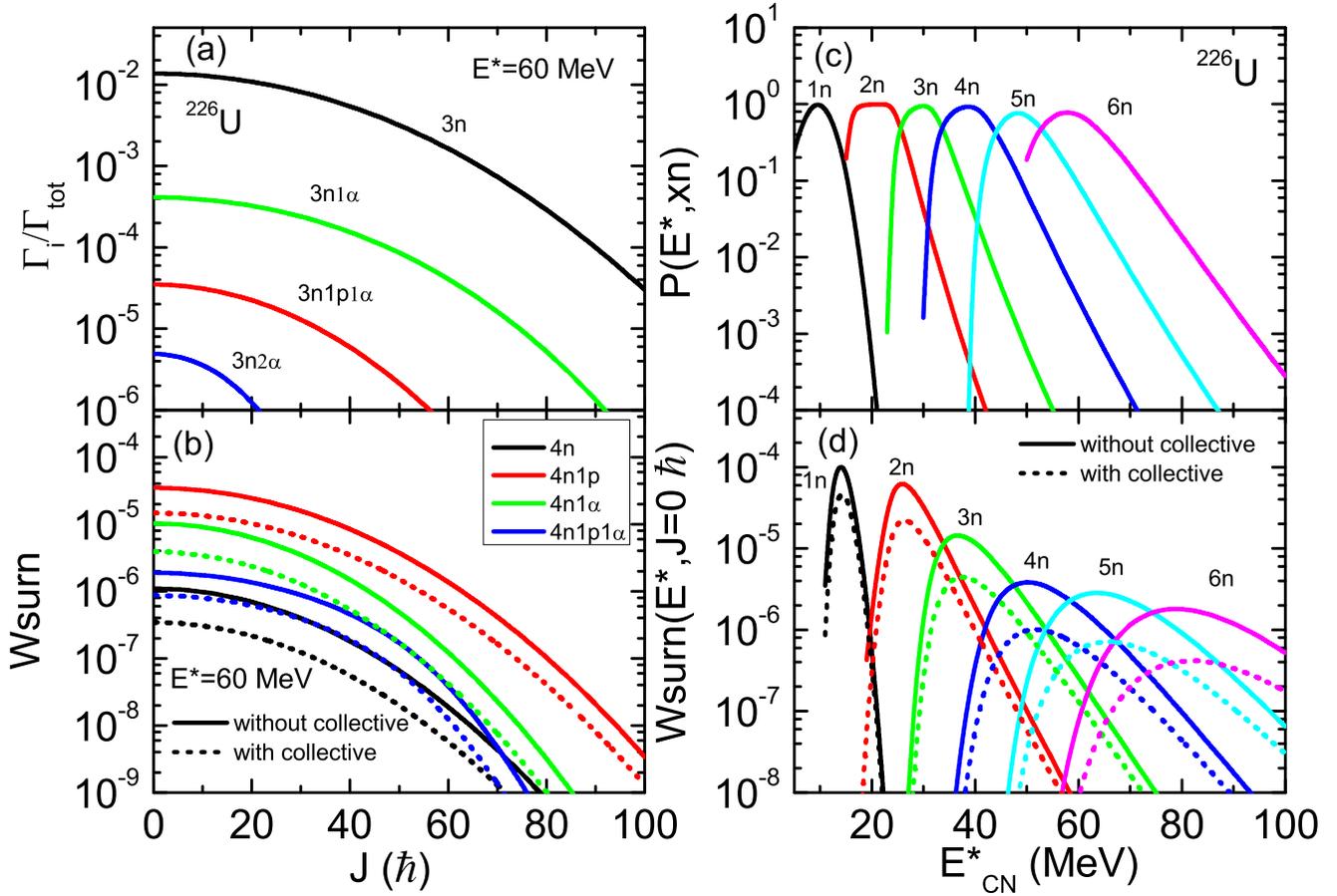}}
\end{center}
\caption{(a) Ratio of partial to total decay width by evaporating different particles and (b)  survival probability as a function of angular momentum.  (c) Excitation energy dependence of realization probabilities and (d) survival probabilities in neutron channels.}
\end{figure}

\begin{figure}
\begin{center}
{\includegraphics*[width=1.\textwidth]{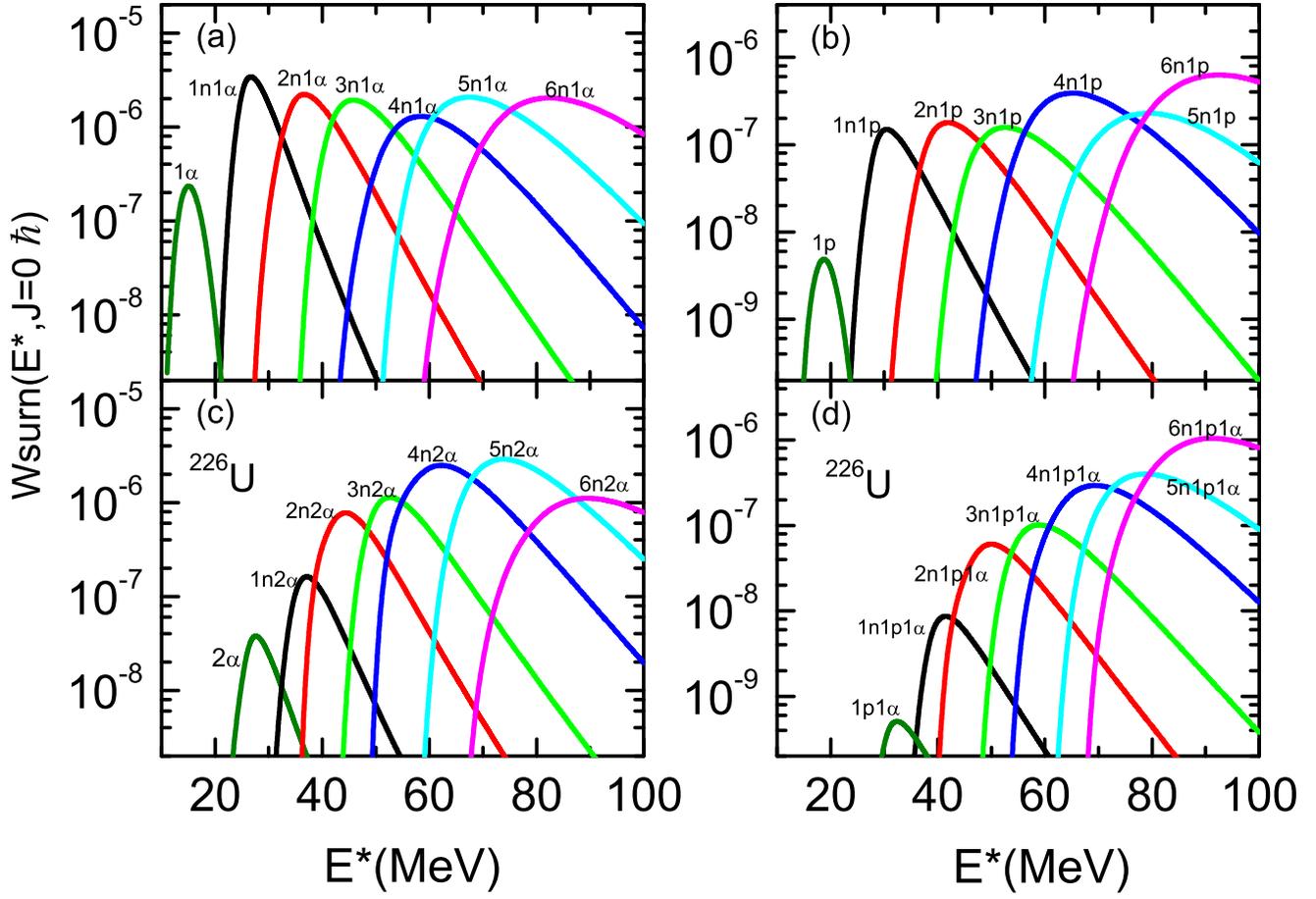}}
\end{center}
\caption{The survival probabilities of proton-rich nucleus $^{226}$U at the angular momentum of $J = 0 \hbar$ as functions of excitation energies in the channels of $1\alpha xn$, $1p xn$, $2\alpha xn$ and $1\alpha 1p xn$ with $x=0-6$.}
\end{figure}

We disintegrate the ingredients in evaluating the survival probability. Shown in Fig. 2 is the comparison of different channels on the ratio of partial to total decay width, the realization and survival probabilities of $^{226}$U. The maximum position of the realization probability in each $xn$ channel is similar and close to unity. However, the survival probabilities decrease with increasing the neutron numbers. The collective enhancement on the level density reduces the survival probability of excited nucleus. The channels with evaporating charged particles are shown in Fig. 3, i.e., $1\alpha xn$, $1p xn$, $2\alpha xn$ and $1\alpha 1p xn$ with $x=0-6$. The maximal probability is different for each channels with evaporating 1$\alpha$, 1$p$, 2$\alpha$ and 1$\alpha$1$p$, e.g., the optimal channels $1n1\alpha$ , $4n1p$, $6n1p$, $4n2\alpha$, $5n2\alpha$ and $6n1p1\alpha$. The maximal cross sections of evaporation residue nuclei with increasing incident energy are contributed from the survival probabilities of compound nucleus for each channel and fusion cross sections of two colliding partners. The channels of evaporating charged particles have comparable values to pure neutron channels. Therefore, it is necessary to include the charged particle evaporation for the excited proton-rich nuclei.

As a test of the approach, we calculated the fusion-evaporation excitation functions in the reaction $^{28}$Si + $^{198}$Pt as shown in Fig. 4. Production cross sections of the residue nuclei from the available experimental data\cite{Ni00} can be nicely reproduced. The solid lines are the total cross sections as the sum of each evaporation channels labeled in the figure. The capture cross sections are calculated by the empirical coupled channel model (EMCC) and we set the fusion probability $P_{CN}\approx 1$. The quasi-fission reactions for the considered system are negligible. It is interest to be noticed that the channels with charged particle evaporation have the larger cross sections than the pure neutron evaporation, i.e., $1\alpha xn$ with $x$=3-6. Because of the smaller separation energy for the proton-rich nucleus, the $\alpha$ particles could be easily emitted from the excited mother nucleus in comparison to protons although the larger Coulomb barrier. The higher fission barrier leads to the larger survival probability, which is favorable to produce the proton-rich nuclei in experiments, even close to the proton drip line.

In summary, a statistical approach based on the Weisskopf evaporation theory has been developed to describe the de-excitation process of highly excited heavy and superheavy nuclei, in particular for the proton-rich nuclei, which is the most important stage in the evaluation of residue cross section. The light charged particle evaporation has been included in the model. The fusion-evaporation excitation functions of the available experimental data could be nicely reproduced. The approach is useful to evaluate the production cross section close to the proton (neutron) drip lines in fusion-evaporation reactions and multi-nucleon transfer reactions, which are particularly interests in the near future experiments at HIRFL (heavy-ion accelerator facility in Lanzhou).

\begin{figure}
\begin{center}
{\includegraphics*[width=1.\textwidth]{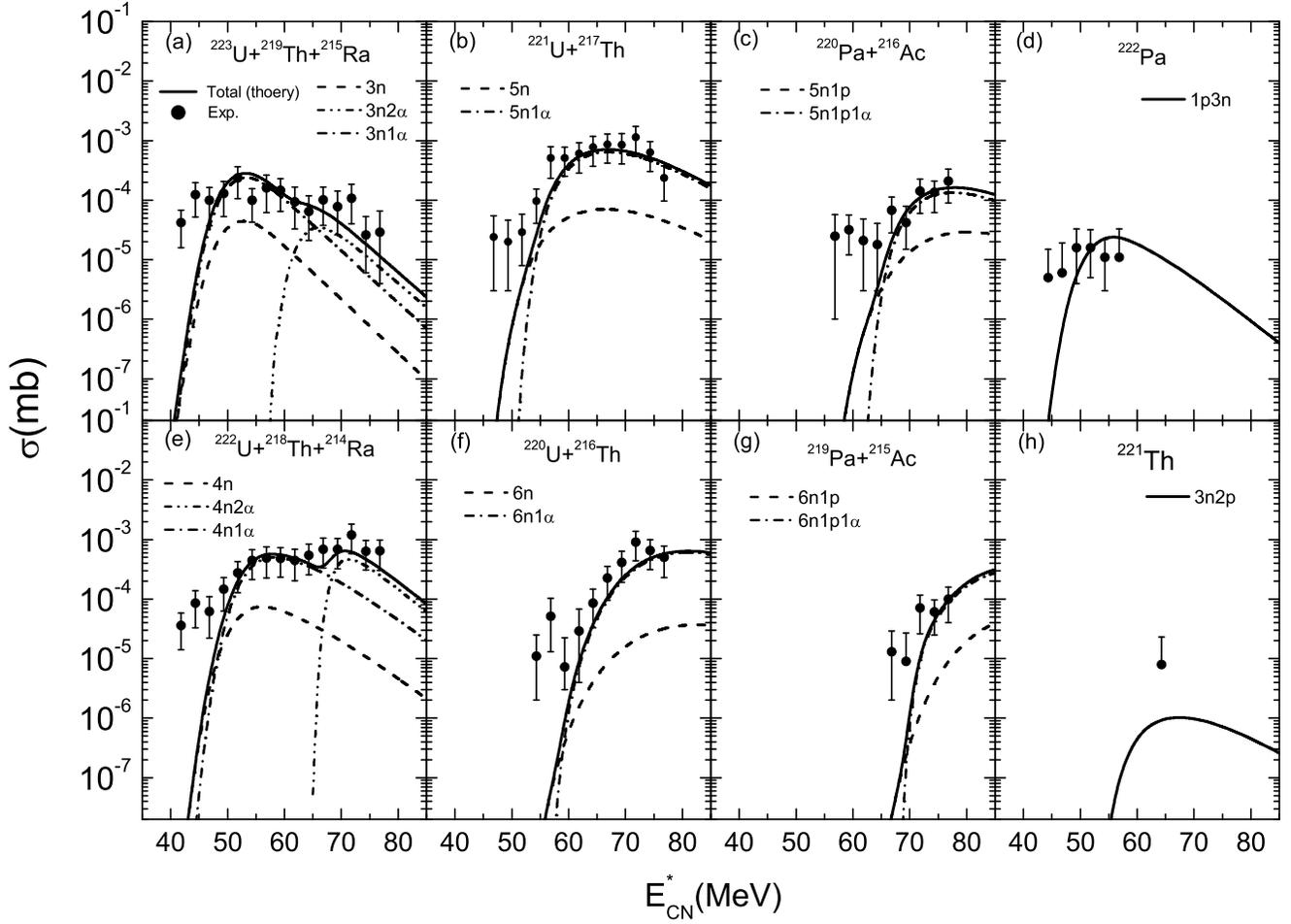}}
\end{center}
\caption{Fusion-evaporation excitation functions for all possible channels in the reaction $^{28}$Si + $^{198}$Pt and compared with the experimental data \cite{Ni00}.}
\end{figure}

\textbf{Acknowledgements}

 This work was supported by the Major State Basic Research Development Program in China under Grant No 2015CB856903, the National Natural Science Foundation of China Projects under Grant Nos 11175218, U1332207, 11475050, 11175074, and the Youth Innovation Promotion Association of Chinese Academy of Sciences.


\begin{thebibliography}{99}

\bibitem{Ho00} S. Hofmann and G. M\"{u}nzenberg,  Rev. Mod. Phys., 72: 733 (2000); G. M\"{u}nzenberg, Nucl. Phys. A, \textbf{944}: 5 (2015)
\bibitem{Mo04} K. Morita et al, J. Phys. Soc. Jpn, \textbf{73}: 2593 (2004); K.  Morita, Nucl. Phys. A, \textbf{944}: 30 (2015)
\bibitem{Og07} Y. Oganessian, J. Phys. G, {\bf 34}: R165 (2007); Y. Oganessian and V. K. Utyonkov, Nucl. Phys. A, {\bf 944} 62: (2015 )
\bibitem{Fe06} Z. Q. Feng, G. M. Jin, F. Fu, and J. Q. Li, Nucl. Phys. A, {\bf 771}: 50 ( 2006)
\bibitem{We37} V. Weisskopf, Phys. Rev., {\bf 52}: 295 (1937)
\bibitem{Za03} V. I. Zagrebaev, Phys. Rev. C, {\bf 67}: 061601 (R) (2003)
\bibitem{Ya13}  R. Yanez et al, Phys. Rev. C  {\bf 88}: 014606 (2013)
\bibitem{Sc91} K. H. Schmidt, W. Morawek, Rep. Prog. Phys., {\bf 54}: 949 (1991)
\bibitem{Fe07} Z. Q. Feng, G. M. Jin, J. Q. Li, and W. Scheid, Phys. Rev. C,  {\bf 76}: 044606 ( 2007)
\bibitem{Fe09}  Z. Q. Feng, G. M. Jin, and J. Q. Li, Nucl. Phys. A, {\bf 816}: 33 (2009)
\bibitem{Ig79} A. V. Ignatyuk, K. K. Istekov, and G. N. Smirenkin, Nucl. Phys., {\bf 29}: 875 (1979)
\bibitem{Ju98} A. R. Junghans, M. de Jong, H. G. Clerc et al, Nucl. Phys. A, {\bf 629}: 635 (1998)
\bibitem{Ja56} J. D. Jackson, Can. J. Phys., {\bf  34}: 767 (1956)
\bibitem{Ni00} K. Nishio, H. Ikezoe, S. Mitsuoka et al, Phys. Rev. C, {\bf 62}: 014602 (2000)

\end{thebibliography}
\end{document}